\documentclass[conference]{IEEEtran}

\ifCLASSOPTIONcompsoc
  \usepackage[nocompress]{cite}
\else
  \usepackage{cite}
\fi

\usepackage[pdftex]{graphicx}
\graphicspath{{iolts2019/figures/}} 
\usepackage{amsmath}
\usepackage{xcolor}
\usepackage[linesnumbered,ruled,lined,commentsnumbered]{algorithm2e}
\usepackage{url}
\usepackage{algpseudocode}
\usepackage[export]{adjustbox}

\usepackage{times}
\usepackage{amssymb}
\usepackage{url}
\usepackage{hyperref}
\usepackage{amsmath}
\usepackage{multirow}
\usepackage{graphicx}
\usepackage{setspace}
\usepackage{wrapfig}
\usepackage{subcaption}
\usepackage{dblfloatfix}    

\makeatletter
\newcommand{\removelatexerror}{\let\@latex@error\@gobble}
\makeatother

\begin{document}
\title{PASCAL: Timing SCA Resistant Design and Verification Flow}

\author{Xinhui Lai$^1$, Maksim Jenihhin$^1$, Jaan Raik$^1$, Kolin Paul$^{1,2}$ \\
	$^1$ Computer Systems, Tallinn University of Technology, Estonia\\
	$^2$ Department of Computer Science \& Engg. Indian Institute of Technology Delhi, India \\
	email:{\{Xinhui.Lai,Maksim.Jenihhin,Jaan.Raik,Kolin.Paul\}@taltech.ee}

}
\onecolumn
\noindent\textcopyright 2019 IEEE. Personal use of this material is permitted. Permission from IEEE must be obtained for all
other uses, in any current or future media, including reprinting/republishing this material for advertising or
promotional purposes, creating new collective works, for resale or redistribution to servers or lists, or reuse
of any copyrighted component of this work in other works. 

\twocolumn
\maketitle
\begin{abstract}

A large  number of crypto accelerators are being deployed  with the widespread adoption of IoT.  It is vitally important that these accelerators and other security hardware IPs are \textit{provably secure}. Security is  an extra functional requirement and hence many security verification  tools are not mature. We propose an approach/flow -- PASCAL --  that works on RTL designs and discovers potential Timing Side Channel Attack (SCA) vulnerabilities in them. Based on information  flow analysis, this is able to identify Timing Disparate Security Paths that could lead to information leakage. This flow also (automatically) eliminates the information leakage caused by the timing channel.  The insertion of a lightweight \textit{Compensator Block}  as balancing or compliance FSM removes the timing channel with minimum modifications to the design with \textbf{no impact} on the clock cycle time or combinational delay of the critical path in the circuit.

\end{abstract}

\IEEEpeerreviewmaketitle

\vspace*{-0.1cm}
\section{Introduction}
\vspace*{-0.1cm}

\label{sec:intro}
Security is not a first class citizen in (hardware) design and  is rarely considered during design space exploration. Bugs or vulnerabilities can originate from design flaws, some of which can be fully eliminated after a complete verification. The goal of the adversary in a security critical application, is to learn information that one has no legitimate access to, e.g. the classified data or secret keys.  Novel attack vectors like side-channel analysis rely on design features, to build efficient exploits that undermine assumptions regarding the accessibility of internal secret information in a computing system.  For example, Timing Driven Attacks exploit timing differences in execution traces as the information flow is via different paths with the same start and end nodes (controllable and observable) to derive the secret information. 
\par Denning et. al. introduced the concept of secure information flow in a computer system whereby it can be shown that no unauthorized flow of information is possible due to control and data flow~\cite{Denning76}.  However, in recent years, side channels or out of band data channels have been exploited to exfiltrate or deduce secret information. Consequently, Information Flow Tracking (IFT) has evolved and has been used as a formal methodology for modeling and reasoning about security properties related to integrity, confidentiality of side channels. The problem becomes more interesting and hard because high-level architecture abstractions are translated into transparent micro architecture implementations. While the hardware behavior in the micro-architecture can cause additional information flows  which can be gainfully exploited to form these side channels.
\par
\begin{figure}[h]
\begin{center}
\includegraphics[scale=0.4]{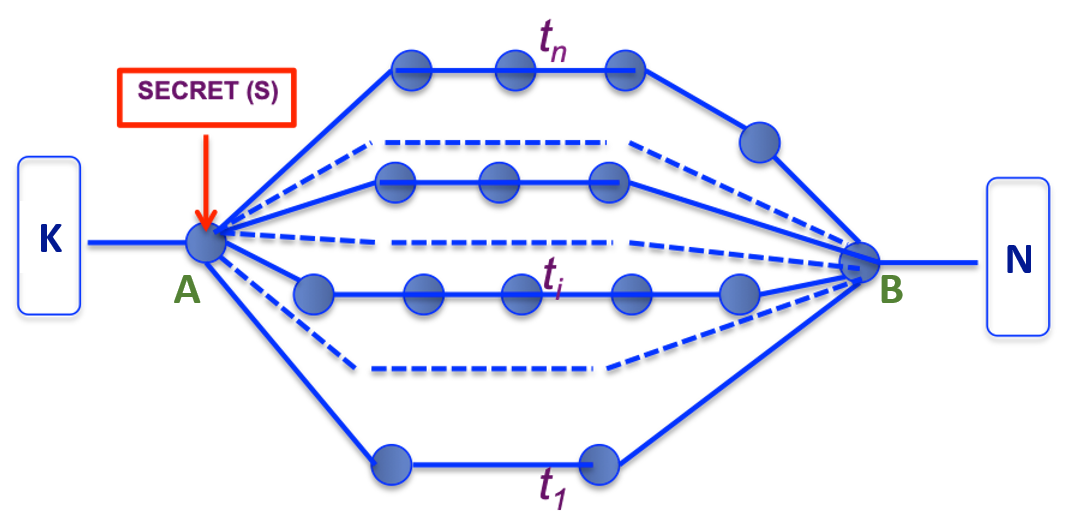}
	\caption{ Timing Disparate Paths }
	\label{fig:timingChannel}
\end{center}	
\end{figure}

 As opposed to physical Side Channels Attacks (SCA) like differential power attacks etc. that require physical access to the computer system, Timing SCA can be launched  (relatively) easily on general purpose compute environments that contain a memory hierarchy or performance enhancing microarchitecture features like speculative execution. The key invariant in  these attacks is that there are different timing paths that provide out of band information.
Security path verification  addresses a specific, important aspect of overall security verification by checking access to the secure data on the hardware to make certain that attackers access the secure (secret) data through illegal logic paths. 
For example, in Figure~\ref{fig:timingChannel}, 
\vspace*{-5pt}
\vspace*{-5pt}there are paths from A to B which is controlled by the node S containing the secret. Tools do Taint Propagation/Taint Analysis, which is a conservative approximation of secure information flow analysis, to find such paths~\cite{Ming2015}. A timing side channel exists, if the contents of S can be derived/deduced by analyzing time of arrival of K at N. We call these two or more paths with unequal transit time as \textbf{Timing Disparate Security Paths}. And these Timing Disparate Security Paths will be potentially vulnerabilities open for a timing side channel exploit.
\par The primary contribution of our work is a secure automated digital design flow -- \textbf{PA}th based \textbf{S}ide \textbf{C}hannel \textbf{A}na\textbf{L}ysis (\textbf{PASCAL}) -- that creates a secure IP core or system-on-chip. The proposed flow starts from the RTL design  and the threat model and uses a state of the art Security Path verification tool to  identify   potential timing side  channel vulnerabilities and proposes a method to remove them by enforcing uniform timing to remove data dependent instruction cycle count variations in the timing side channels.
\par The remainder of the paper is organized as follows. Section~\ref{sec:related} summarizes the state of the art in this area. The next section details the approach used and presents the key algorithmic contribution in this work for identifying Timing Disparate Security Paths while Section~\ref{sec:mitigation} presents a lightweight method for  Timing SCA Resistant Design using the results of the method proposed in the previous section.  Section~\ref{sec:results} describes the implementation results on a widely used crypto core and also demonstrates the efficacy of the proposed mitigation method. Section~\ref{sec:conc} summarises the contributions of the paper and provides directions for future work.

\vspace*{-0.1cm}
\section{Related Work}
\vspace*{-0.1cm}

\label{sec:related}

Timing side channel attacks
are  known to be a hard and a very important problem in modern systems.  They have been used to extract cryptographic secrets
from running systems.
 Even differential privacy systems are not immune to these attacks. 
And these are possible using both remote
and local adversaries
. Koeune et. al present an indepth tutorial on Side Channel Attacks~\cite{Koeune2005}.

\par A popular approach for defending against both local and
remote timing attacks is to ensure that the low-level instruction
sequence does not contain instructions whose performance
depends on secret information. This can be enforced by
manually re-writing the code, as was done in OpenSSL or by
changing the compiler to ensure that the generated code has
this property~\cite{CoppensVBS09}.
\par While methods for high performance design or low power are available, design for security is still adhoc. Only recently, systematic methods  support design for
trust and security have been described in literature~\cite{Tiri2005}.
Menichelli et. al present  an exploration approach centered on high level simulation based on SystemC to  suggest improvements in the knowledge and  identification of the weaknesses in cryptographic algorithm implementations~\cite{Menichelli2008}. 
Ardeshiricham et. al. have proposed an information flow based method for secure hardware design~\cite{Ardeshiricham2017} by analyzing all logical code flows of the RTL code. In contrast VeriCoq-IFT converts designs from HDL to Coq to analyze a formal security properties~\cite{Bidmeshki2015}. SecVerilog requires explicit annotating each variable in the design  with a security label --- this is similar to using a type system to track information flow in the code~\cite{Zhang2015}. Deng et. al. have proposed a  Computation Tree Logic to model execution paths of the processor cache logic and derive formulas for paths that can lead to timing side-channel vulnerabilities~\cite{Deng2018}. 

\par Most of  the mitigation techniques that have been proposed  try to remove data dependent instruction cycle count variations by balancing timing or do a power flattening to remove power peaks/anomalies~\cite{Oswald2005}. In some cases,  Pipeline randomization for power and timing is also attempted. Alternatively, packet route randomisation as a mechanism to increase NoC resilience has also been proposed~\cite{IndrusiakHS16}. Recently, Jiang et. al. have proposed a high-level synthesis (HLS) infrastructure that incorporates static information flow analysis to remove  timing channels in a verifiable manner  on HLS-generated hardware accelerators~\cite{Jiang2018}.
\par The methodology proposed in this paper is based on a formal method that can identify \textbf{all} Timing Disparate Security Paths at RT level and improve the state of the art is a simple mitigation scheme for potential SCA vulnerable timing channels.
\par In the next section, we discuss the proposed method for discovering Timing Disparate Security Paths in RTL designs.
\vspace*{-0.1cm}
\section{Methodology}
\vspace*{-0.1cm}

\label{sec:method}
\par Hardware implementations of encryption algorithms are being increasingly used as hardware is regarded as more effective root of trust. RSA is a asymmetric cryptographic algorithm and has been shown to be vulnerable to Timing SCAs and mitigation techniques have also been proposed. However, the major focus continues to rely on verifying the correctness of encryption algorithms and their implementation in software and hardware. We present an approach based on RT level analysis that allows a precise understanding of possible flows for side channels based on timing. The methodology relies on a formal analysis tool Cadence JasperGold Security Path Verification App (JG SPV)~\cite{cadenceJG}. The original objective of the tool is for security verification by checking access or leak of the secure data on the hardware to make certain that the attacker cannot breach the authentication logic and seek the secure data through illegal paths.

\par Based on a formal method of path sensitization from the secret information to the output observable points, we propose a method that can detect possible Timing-Disparate Paths in RTL designs which could be exploited as Timing Side Channel(s). As a result of this analysis, a simple and effective retiming of Timing SCA sensitive paths is proposed to make the design immune for the threats under the chosen threat model. We illustrate this on a standard RSA RTL Verilog code.
\vspace*{-5pt}
\begin{algorithm}[!]
	\DontPrintSemicolon
	\caption{Example: RSA Modulus Code\label{algo:rsa}}
	\KwIn{$C_{m},P_{n}$; // C is the \textit{m} bits cipher text, P is the \textit{n} bits private key}
	\KwOut{$O_{m}$; // O is the \textit{m} bits output plain text}
	$R_0 \leftarrow Montgomery(C_{m})$ and $R_1 \leftarrow Montgomery(1))$ \;
	$j \leftarrow 0$ \;
	\While{$j \le n-1$}
	{	
		$ R_0 \leftarrow Montgomery\_Reduction(R_1*R_1)$ \;
		\If{$P[j]$}
		{
			$R_0 \leftarrow Montgomery\_Reduction(R_0 * R_1)$ \;
		}
		$O_{m} \leftarrow Montgomery^{-1}(R_0)$ \;
	}
\end{algorithm}
\vspace*{-5pt}
\par The decryption of the RSA modulus in \textbf{Algorithm~\ref{algo:rsa}} uses Montgomery modular multiplication with square-and-multiply algorithm. Here we did not mention the details about how to choose the key or how the Montegomery algorithm works but focus on explaining the unintended timing channels in RSA which can be used by attacker to reverse the private key. In \textbf{Algorithm~\ref{algo:rsa}}, n, the bit number of \textbf{$P_{n}$}, decides total loop times while value of single bit of \textbf{$P_{n}$}: P[j] determine the operations for each single loop -- only when P[j] equal to 1, statements at Line 5,6,7 will be executed while P[j] equal to 0 will not. For the decryption of RSA, the total operations need to be executed might be different with different private key duing to the above reasons. Assuming the time for single bit P[j] is $t_{\textbf{P[j]}}$, the final execution time will be $t_{total} = \sum_{j=0}^{n}t_{\textbf{P[j]}}$. Thus keys with different number of '1s' will cause the different execution time. This will open a timing side channel for the attackers.
\par
\begin{figure}[h]
		\includegraphics[scale=0.5]{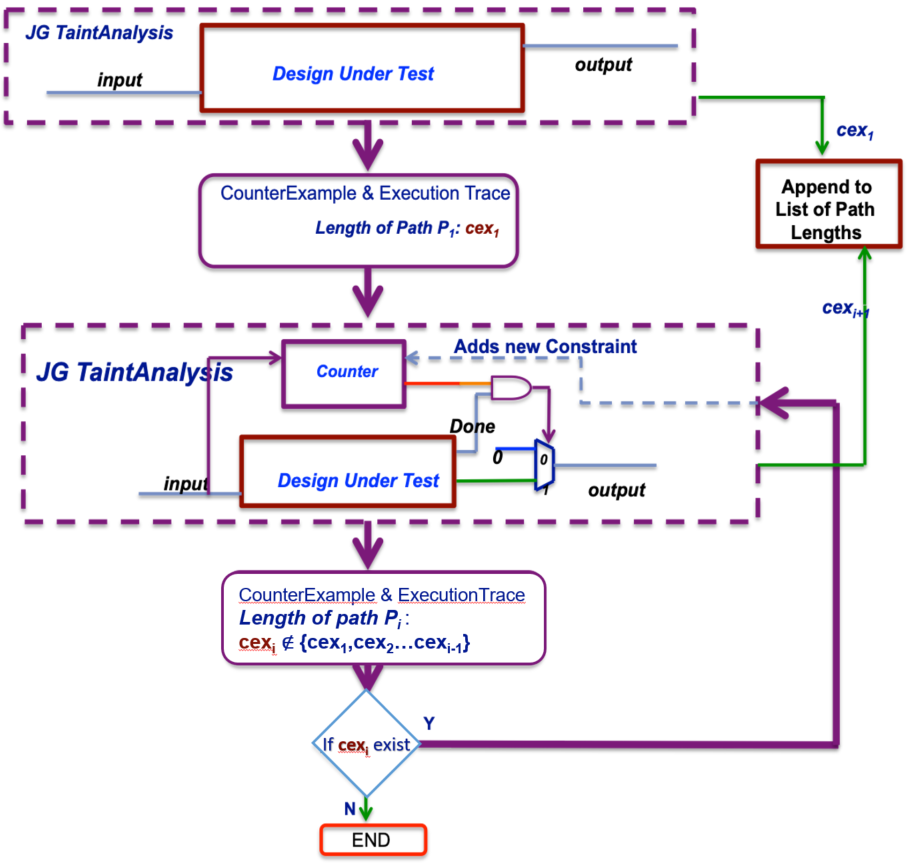}
		\caption{\textbf{PASCAL}: Graphical Representation}
		\label{fig:Algo}
	\end{figure}

\begin{figure}[!h]
    \includegraphics[scale=0.167]{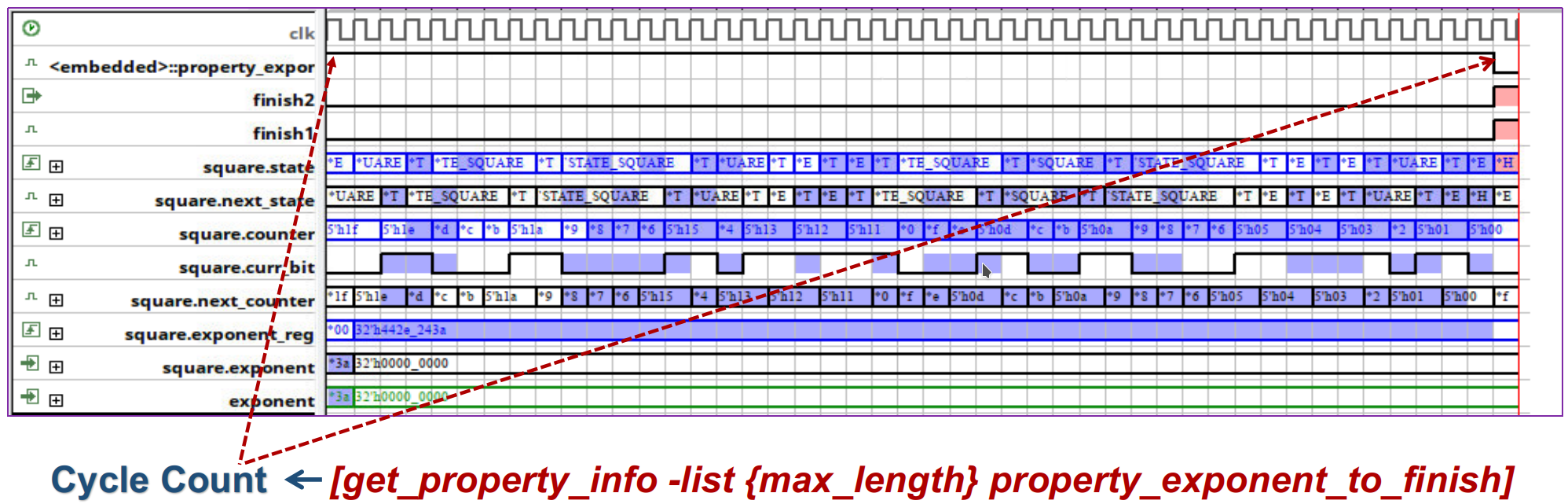}
    \caption{Counter Example and Execution Trace}
    \label{fig:my_label}
\end{figure}

\begin{figure}[!h]
	    \includegraphics[height=3cm]{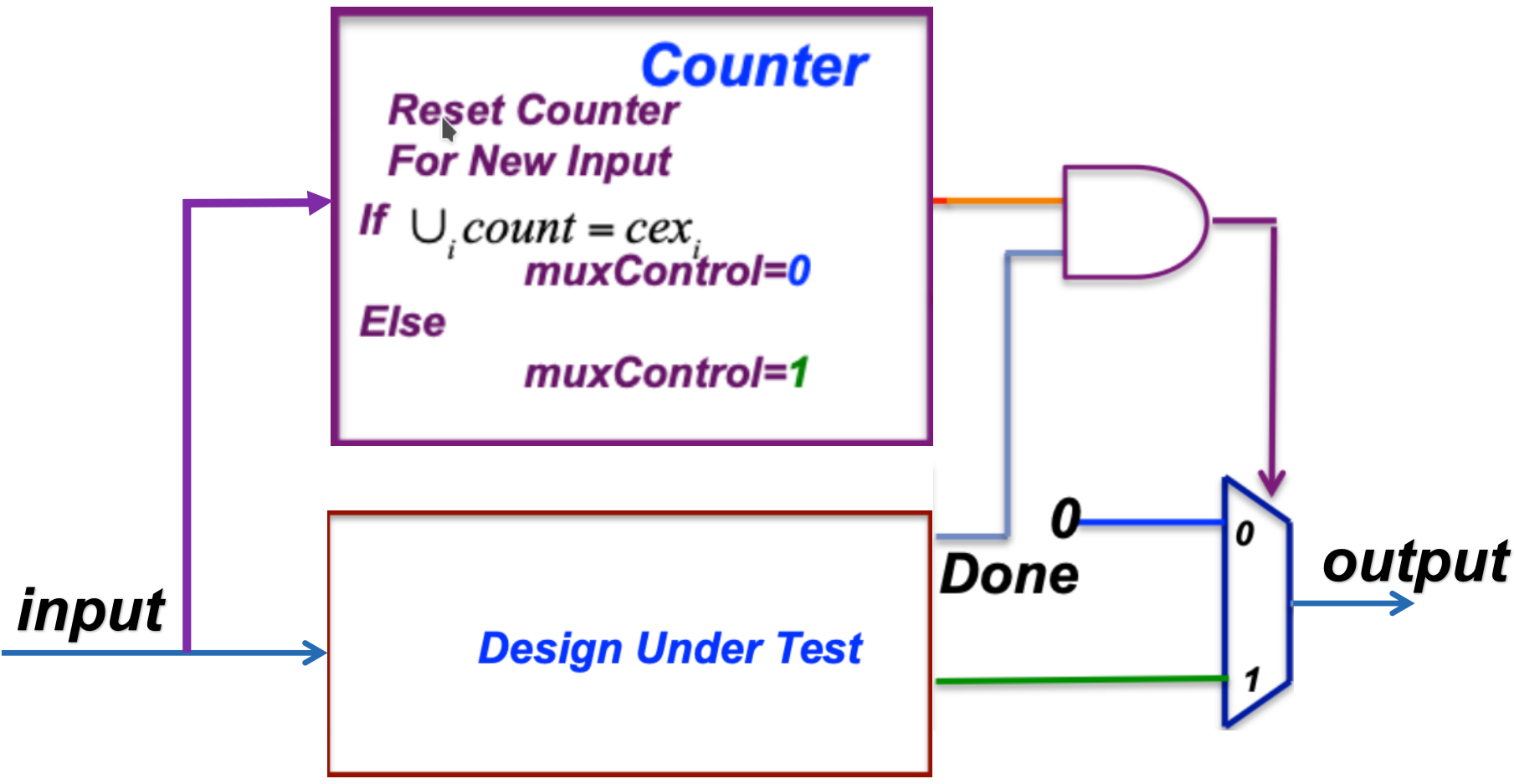}
	    \caption{\textbf{Modified DUV}}
	    \label{fig:mDUV}
	\end{figure}

 For this Timing SCAs, the \textbf{PASCAL} is shown in figure~\ref{fig:Algo}.
\vspace*{-5pt}
\vspace*{-5pt}
Firstly, we use JG SPV to analyze if there is one or several paths, from a variable deemed to be secure and unobservable to the output, exist. JG SPV uses a special path sensitization technology implying taint analysis to find if private key \textbf{P} can be propagated to the output \textbf{O}. Then if the path exist, JG SPV will give a counterexample along with an execution trace detailing: the exact number of clock cycles(say X). As shown in the figure~\ref{fig:my_label}, the example shows waveforms of related signals along the path. We use the command "\textit{[get\_property\_info -list\{max\_length\} property\_exponent\_to\_finish]}" to get the total execution time(clock based) of an exist path for this specified secure signal pair. Here it needs 44 clock cycles(additional 2 clock cycles are for setting up)to propagate.
\vspace*{-5pt}
After that, JG SPV will be used to find another functional path (if it exists) from \textbf{P} to the output \textbf{O} with a time length different from X cycles. This is achieved by invoking JG SPV on a modified design, shown in Figure ~\ref{fig:mDUV}. A counter is added which drives the multiplexer to select the situation where the DUV  both finish the decryption \textit{AND} also the length of the execution trace Y is not equal to X. If JG SPV finds another path with an execution trace length not equal to X and Y, it is added to the  \textit{Union Clause} of the  multiplexer select 
condition and the process is repeated until they find all the timing classes. 
\newcommand{\myalgorithm}{%
\begingroup
\removelatexerror
\begin{algorithm}[!hbt]
	\caption{\textbf{PASCAL: Algorithm}}
		\label{fig:algomethod}
		\KwIn{Design Under Verification (DUV)}
		\KwOut{List of timing disparate Paths (P)  }
		cex$_1$ = Execute JG SPV with DUV\;
		Append Path P$_1$ to P\;
		$i=1$\;
		\While{ cex$_i!=0$}{
			Add cex$_i$ as new constraint \;
			Initialize Counter\;
			cex$_{i+1}$ $\leftarrow$ Execute JG SPV with modified DUV\;
			Append Path P$_i$ to P\;
			$i \leftarrow i+1$\;
		}
	\end{algorithm}	
	\endgroup}
\vspace*{-0.1cm}
\section{Timing SCA Secure Design Flow }
\vspace*{-0.1cm}

\label{sec:mitigation}
We also propose a method that aims to achieve timing-sensitive noninterference for the
synthesized design, via which it is ensured  that confidential or secret values
cannot be revealed by the observing/measuring the timing of events at observable ports. An intuitive method to remove this Timing SCA vulnerability is to insert additional registers in the faster paths using path-balanced scheduling~\cite{Peter2016}. 
However, as shown in Figure~\ref{fig:timingChannel}, there could be many paths $t_1,t_2,\cdots,t_n$in the same basic block. Assume without loss of generality, that there are $n/2$ paths each differing by one cycle. Hence a path balanced scheduling synthesis procedure would insert $1+2+3+\cdots+n/2$ or $\mathcal{O}(n^2)$ registers.

\par The method we propose is shown in \textbf{Algorithm~\ref{algo:balance}}. Since Timing Disparate Security Paths result in a Timing SCA vulnerability only if they are observable at user interfaces (output ports), it can relax the constrains in the path balanced scheduling approach and enforce indistinguishable timing behaviors at the observable points in the design. 
Clearly, for the basic block or the core to be timing insensitive at the observable points, the output should be observable modulo $t_{max}$ cycles where $t_{max}=maximum(t_1,t_2,\cdots,t_n)$. We enable the output port/interface every $t_{max}$ cycles using a counter and an AND gate. This small additional circuitry acts as the \textbf{Compensator} or \textbf{balancing/compliance FSM} and provides the (read) enable / data ready signal for observable interface. 
\par This therefore, leads to a very simple synthesis technique for ensuring a path balanced design with a single lightweight \textbf{Compensator Block} at the observable points of interest in the design. The additional circuit has a very small overhead counter which counts upto $t_{m}$ to generate the control input for the AND gate which provides the enable signal to the observable register. The counter is reset every time a new input enters the  basic block. This additional logic incurs no penalty in the critical path of the system and avoids resource duplication since
\begin{algorithm}
	\caption{Timing Channel Removal\label{algo:balance}}
	\KwIn{Design Under Verification (DUV)}
	\KwOut{Secure Design Under Test (sDUV)}
	$P \leftarrow $ PACSCAL(DUV) \;
	t$_{m}$ = findMaxExecutionLength(P)\;
	\textit{Ccompensator Logic Block }$\leftarrow$ Counter($t_m$) + ResetLogic\;
	sDUV $\leftarrow$  \textit{Compensator Logic  Block} + DUV
\end{algorithm}
it has a uniform counter where the results from the different Timing Disparate Security Paths are  delivered to the observable interface with the same latency.
\vspace*{-0.1cm}
\section{Results}
\vspace*{-0.1cm}

\begin{figure}[hbt]
	\includegraphics[scale=0.6]{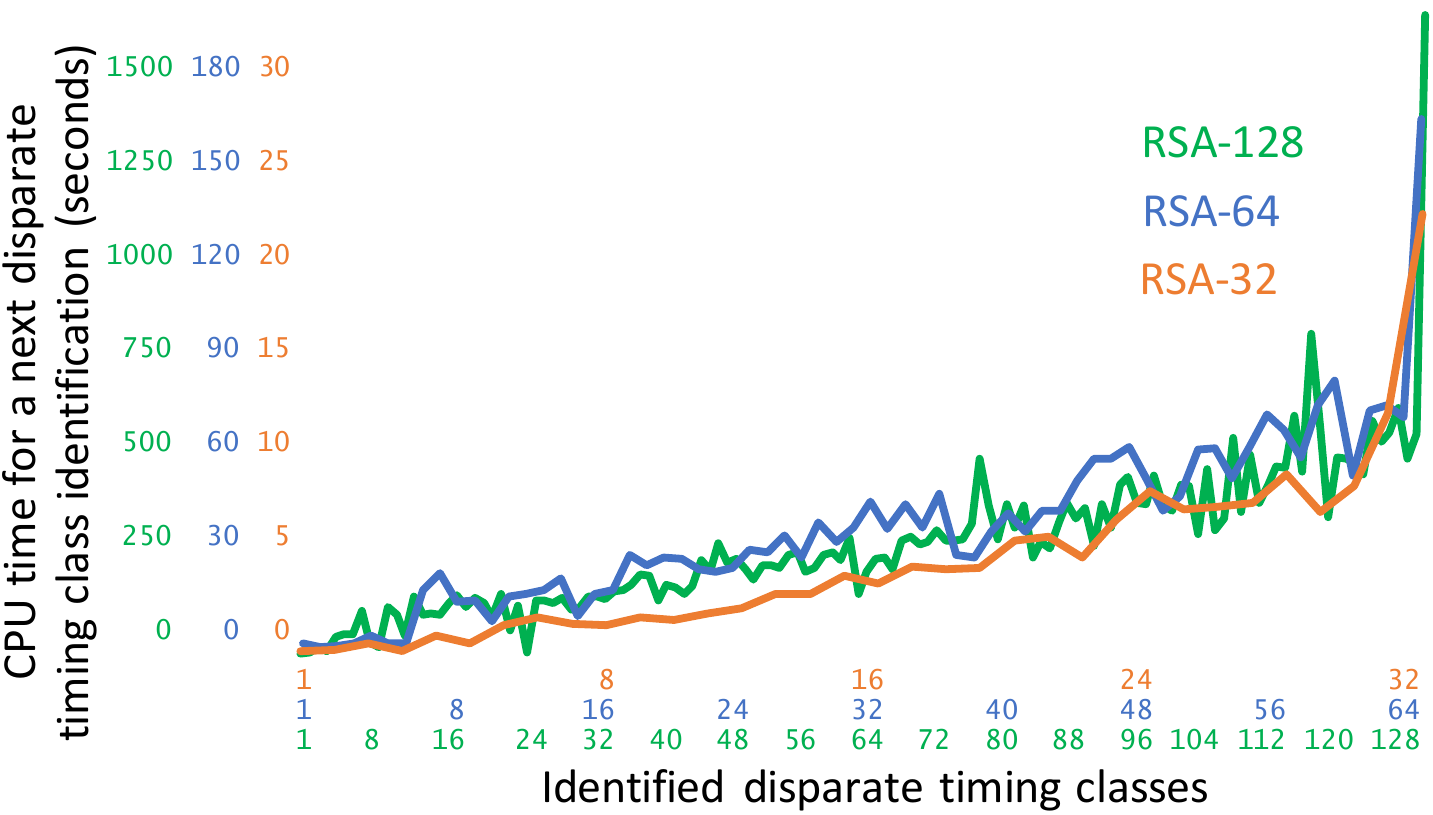}
	\caption{Normalized Execution Times}
	\label{fig:time}
\end{figure}

\label{sec:results}
The Montgomery modular multiplication with square-and-multiply algorithm based RSA cryptographic RTL implementation is vulnerable to timing SCA. This is because for different keys the time differences are dependent on the number of '1s' in the key as explained in the Algorithm~\ref{algo:rsa}. In figure~\ref{fig:time}, the time required to generate the timing disparate classes for 32-bits RSA, 64-bits RSA and 128-bits RSA are shown. For different RSA, the time needed to identify timing cases are varies: for the initial few time classes, they are obtained quickly while for the last few time classes they need a very long time. 
\par Our method can correctly identifies all timing classes using formal methods. i.e. for the 32-bits RSA verilog implementation, it identifies all the timing classes with cycle times from 33 to 64. As for the mitigation method mentioned in Figure~\ref{fig:mDUV}. Since the counter need to count to 64, we only need a 7 bits counter which incurs an approximate area penalty of 7 flops. In contrast, the path-balanced scheduling strategy would require about 512 flip flops. Clearly, with a 64-bit RSA, the savings are more significant. As mentioned earlier, the Compliance State Machine is not in the critical path and incurs no penalty in the operational speed of the circuit. 

\vspace*{-0.1cm}
\section{Conclusion}
\vspace*{-0.1cm}

\label{sec:conc}
Significant numbers of hardware IPs or crypto accelerators are being deployed  with the widespread adoption of IoT. It is vitally important that these IPs are \textit{provably secure}. We have proposed a novel approach to discover  timing SCA vulnerabilities that (can)exist in designs. This flow also (automatically) eliminates the information leakage caused by the timing channel. The insertion of a lightweight Compensator Block removes the timing channel with minimum modifications to the design with \textbf{no impact} on the clock cycle time or combinational delay of the critical path in the circuit. For the future work, multiple secrets in design or multiple public interfaces will be studied. And we will also integrate this framework to High Level Synthesis flow so that more accurate estimates of area can be obtained.
\vspace*{-0.1cm}
\section*{Acknowledgements}
\vspace*{-0.1cm}

\par {\small This research was supported in part by projects H2020 MSCA ITN RESCUE funded from the EU H2020 programme under the MSC grant agreement No.722325, by the Estonian Ministry of Education and Research institutional research grant no. IUT19-1 and by European Union through the European Structural and Regional Development Funds.}
\bibliographystyle{IEEEtran} 
\bibliography{IEEEabrv,iolts2019}

\begin{thebibliography}{10}
\providecommand{\url}[1]{#1}
\csname url@samestyle\endcsname
\providecommand{\newblock}{\relax}
\providecommand{\bibinfo}[2]{#2}
\providecommand{\BIBentrySTDinterwordspacing}{\spaceskip=0pt\relax}
\providecommand{\BIBentryALTinterwordstretchfactor}{4}
\providecommand{\BIBentryALTinterwordspacing}{\spaceskip=\fontdimen2\font plus
\BIBentryALTinterwordstretchfactor\fontdimen3\font minus
  \fontdimen4\font\relax}
\providecommand{\BIBforeignlanguage}[2]{{%
\expandafter\ifx\csname l@#1\endcsname\relax
\typeout{** WARNING: IEEEtran.bst: No hyphenation pattern has been}%
\typeout{** loaded for the language `#1'. Using the pattern for}%
\typeout{** the default language instead.}%
\else
\language=\csname l@#1\endcsname
\fi
#2}}
\providecommand{\BIBdecl}{\relax}
\BIBdecl

\bibitem{Denning76}
D.~E. Denning, ``A lattice model of secure information flow,'' \emph{Commun.
  {ACM}}, vol.~19, no.~5, pp. 236--243, 1976.

\bibitem{Ming2015}
J.~Ming, D.~Wu, G.~Xiao, J.~Wang, and P.~Liu, ``Taintpipe: Pipelined symbolic
  taint analysis,'' in \emph{24th {USENIX} Security Symposium ({USENIX}
  Security 15)}.\hskip 1em plus 0.5em minus 0.4em\relax Washington, D.C.:
  {USENIX} Association, 2015, pp. 65--80.

\bibitem{Koeune2005}
F.~Koeune and F.-X. Standaert, ``Foundations of security analysis and design
  iii,'' A.~Aldini, R.~Gorrieri, and F.~Martinelli, Eds.\hskip 1em plus 0.5em
  minus 0.4em\relax Berlin, Heidelberg: Springer-Verlag, 2005, ch. A Tutorial
  on Physical Security and Side-channel Attacks, pp. 78--108.

\bibitem{CoppensVBS09}
B.~Coppens, I.~Verbauwhede, K.~D. Bosschere, and B.~D. Sutter, ``Practical
  mitigations for timing-based side-channel attacks on modern x86 processors,''
  in \emph{30th {IEEE} Symposium on Security and Privacy (S{\&}P 2009), 17-20
  May 2009, Oakland, California, {USA}}, 2009, pp. 45--60.

\bibitem{Tiri2005}
K.~{Tiri} and I.~{Verbauwhede}, ``A vlsi design flow for secure side-channel
  attack resistant ics,'' in \emph{Design, Automation and Test in Europe},
  March 2005, pp. 58--63 Vol. 3.

\bibitem{Menichelli2008}
F.~{Menichelli}, R.~{Menicocci}, M.~{Olivieri}, and A.~{Trifiletti},
  ``High-level side-channel attack modeling and simulation for
  security-critical systems on chips,'' \emph{IEEE Transactions on Dependable
  and Secure Computing}, vol.~5, no.~3, pp. 164--176, July 2008.

\bibitem{Ardeshiricham2017}
A.~Ardeshiricham, W.~Hu, J.~Marxen, and R.~Kastner, ``Register transfer level
  information flow tracking for provably secure hardware design,'' in
  \emph{Proceedings of the Conference on Design, Automation \& Test in Europe},
  ser. DATE '17.\hskip 1em plus 0.5em minus 0.4em\relax 3001 Leuven, Belgium,
  Belgium: European Design and Automation Association, 2017, pp. 1695--1700.

\bibitem{Bidmeshki2015}
M.~Bidmeshki and Y.~Makris, ``Toward automatic proof generation for information
  flow policies in third-party hardware ip,'' in \emph{2015 IEEE International
  Symposium on Hardware Oriented Security and Trust (HOST)}, vol.~00, May 2015,
  pp. 163--168.

\bibitem{Zhang2015}
D.~Zhang, Y.~Wang, G.~E. Suh, and A.~C. Myers, ``A hardware design language for
  timing-sensitive information-flow security,'' \emph{SIGPLAN Not.}, vol.~50,
  no.~4, pp. 503--516, Mar. 2015.

\bibitem{Deng2018}
S.~Deng, W.~Xiong, and J.~Szefer, ``Cache timing side-channel vulnerability
  checking with computation tree logic,'' in \emph{Proceedings of the 7th
  International Workshop on Hardware and Architectural Support for Security and
  Privacy}, ser. HASP '18.\hskip 1em plus 0.5em minus 0.4em\relax New York, NY,
  USA: ACM, 2018, pp. 2:1--2:8.

\bibitem{Oswald2005}
E.~Oswald, S.~Mangard, N.~Pramstaller, and V.~Rijmen, ``A side-channel analysis
  resistant description of the aes s-box,'' in \emph{Fast Software Encryption},
  H.~Gilbert and H.~Handschuh, Eds.\hskip 1em plus 0.5em minus 0.4em\relax
  Berlin, Heidelberg: Springer Berlin Heidelberg, 2005, pp. 413--423.

\bibitem{IndrusiakHS16}
L.~S. Indrusiak, J.~Harbin, and M.~J. Sep{\'{u}}lveda, ``Side-channel attack
  resilience through route randomisation in secure real-time
  networks-on-chip,'' \emph{CoRR}, vol. abs/1607.03450, 2016.

\bibitem{Jiang2018}
Z.~Jiang, S.~Dai, G.~E. Suh, and Z.~Zhang, ``High-level synthesis with
  timing-sensitive information flow enforcement,'' in \emph{Proceedings of the
  International Conference on Computer-Aided Design}, ser. ICCAD '18.\hskip 1em
  plus 0.5em minus 0.4em\relax New York, NY, USA: ACM, 2018, pp. 88:1--88:8.

\bibitem{cadenceJG}
\BIBentryALTinterwordspacing
{JasperGold Security Path Verification App}. [Online]. Available:
  \url{https://www.cadence.com/content/cadence-www/global/en_US/home/tools/system-design-and-verification/formal-and-static-verification/jasper-gold-verification-platform.html}
\BIBentrySTDinterwordspacing

\bibitem{Peter2016}
S.~Peter and T.~Givargis, ``Towards a timing attack aware high-level synthesis
  of integrated circuits,'' in \emph{34th {IEEE} International Conference on
  Computer Design, {ICCD} 2016, Scottsdale, AZ, USA, October 2-5, 2016}, 2016,
  pp. 452--455.

\end{thebibliography}

\end{document}